\documentclass[a4paper,12pt]{article}
\usepackage{jheppub}

\usepackage{lineno}
\usepackage{epsfig}
\usepackage{graphicx}
\usepackage[english,activeacute]{babel}
\usepackage{mathtools}
\usepackage[italic]{hepnames}
\usepackage[scr=boondox,cal=esstix]{mathalpha}
\usepackage{tikz}
\usepackage{tikzscale}
\usetikzlibrary{positioning}
\usetikzlibrary{arrows}
\usetikzlibrary{decorations.pathreplacing,decorations.markings}
            \tikzset{arrow data/.style 2 args={%
            decoration={%
            markings,
            mark=at position #1 with \arrow{#2}},
            postaction=decorate}
            }%
\usetikzlibrary{intersections, backgrounds} % Para las intersecciones de líneas.
\usetikzlibrary{calc}

\usepackage[export]{adjustbox} % Para reescalar subfigures.

\title{\boldmath Further Remarks on the Entanglement Entropy of Hopf links}

 \author[a]{C. J. Ram\'{\i}rez-Valdez}
 \author[b]{H. Garc\'{\i}a-Compe\'an}
 \author[c]{J. de-la-Cruz-Moreno}

 \affiliation[a]{Theoretical Particle Physics and
Cosmology Group, \\
Department of Physics\\
King's College London, Strand, London, WC2R 2LS, U.K.}
 \affiliation[b]{Departamento de F\'{\i}sica,\\
 Centro de Investigaci\'on y de Estudios Avanzados del Instituto Polit\'ecnico Nacional,\\
 P.O. box 14-740, C.P. 07000, Ciudad de M\'exico, Mexico}
 \affiliation[c]
 {Leinweber Center for Theoretical Physics, \\
 Randall Laboratory of Physics\\
  The University of Michigan, Ann Arbor, MI 48109-1040}

\emailAdd{carlos\textunderscore jonathan.ramirez\textunderscore valdez@kcl.ac.uk}
\emailAdd{hugo.compean@cinvestav.mx}
\emailAdd{josedcm@umich.edu}

\abstract{We study the connected sum of Hopf links in $S^3$. Particularly, we compute the entanglement entropy (EE) as a function of the number of link components. We find evidence of lower and upper bounds for the entanglement entropy. We show that the SU$(2)$ theory exhibits sensitivity to the parity of links. We also find evidence suggesting the existence of a well-defined limit of the large number of link components.}

\begin{document}

\hfill LCTP-24-15, KCL-PH-TH/2024-47

\maketitle
\flushbottom

\section{Introduction} 

Quantum entanglement is one of the most puzzling properties of quantum mechanics. It has been understood as the main property that distinguishes between the classical and the quantum regimes. There are various ways to study this phenomenon, from properties like separability of the density matrices for a system which consists of two subsystems \cite{Peres:1996dw}  to properties like the entanglement entropy (EE). The entanglement is a quantum property which measures the amount of correlation between the different subsystems conforming a system.  It is relevant in condensed matter physics, see for instance, \cite{Amico:2007ag,Laflorencie:2015eck}.  The EE is easily defined and computed in quantum mechanics, but in general it is difficult to compute in quantum field theories (QFT's). Thus EE is able to measure the entanglement between parts of a system in theories with many degrees of freedom.  Entanglement entropy is a versatile concept relevant in different fields as in quantum field theories \cite{Calabrese:2004eu,Casini:2009sr}, in holography and quantum gravity, see for instance, \cite{Faulkner:2022mlp,Nishioka:2018khk}. Recently EE has been also experimentally measured in photonic systems \cite{nature}. 

For low dimensional theories it is expected to carry out the computations of EE in a simpler way. The Abelian and Non-Abelian Chern-Simons (CS) gauge theory in three dimensions is one of the theories of interest for describing low-dimensional systems in condensed matter. This theory is the prototype of a theory without dynamical degrees of freedom and with correlation functions of Wilson line operators being topological invariants. It is also formulated as an important example of the axiomatic approach to topological field theories \cite{{atiyah1989},Atiyah:1990dn}. Correlation functions of Wilson line operators of CS theory are shown to be knot and link invariants, being the Jones polynomial the simplest invariant of them \cite{Witten:1988hf}, for a review see \cite{Hu:2001ue,Marino:2004uf}.

Regarding the applications of Non-Abelian CS theory in condensed matter, the FQHE is described in terms of certain quantum states (in a Hilbert space) called non-Abelions \cite{Moore:1991ks}, which are conformal blocks of a certain rational conformal field theory in two dimensions. Anyons were used in quantum computation in \cite{Kitaev:1997wr,Aharonov:2006yxu,Nayak:2008zza}. For systems in two dimensions with topological order, its ground state degeneracy is described in general terms with an EE called topological entropy \cite{Kitaev:2005dm,Levin:2006zz,Affleck:1991tk}. The EE has been computed also in the context of CS theories \cite{Dong:2008ft,Salton:2016qpp}. Moreover, for minimal models of non-oriented unitary rational conformal field theories, it is possible to define a crosscap entropy which possesses many similar characteristics than the one of the topological entropy \cite{Garcia-Compean:2018ury}. In these non-oriented systems the left-right entanglement entropy \cite{PandoZayas:2014wsa,PandoZayas:2016cmz} enters in an interesting way. 

One of the most important aspects in quantum computation is to understand how the EE depends on the global (or topological) properties of the underlying spacetime where the theory is defined. That is, how the global properties enter in the understanding of the entanglement of the system. Another problem is to understand the entanglement in multi-partite systems, i.e., in systems of many subsystems and understand how the many different parts are entangled. These are old problems and they might be related. The topological quantum field theories are the arena where both questions may be addressed. In particular, in \cite{Balasubramanian:2016sro} it was shown how the entanglement entropy can be implemented explicitly in Abelian and Non-Abelian CS gauge theory. In that article, the EE was computed for Hopf links and other links. Motivated by \cite{Balasubramanian:2016sro}, many works expanded these results in various directions \cite{Tan:2017ghv,Dwivedi:2017rnj,Balasubramanian:2018por,Hung:2018rhg,Melnikov:2017bjb,Melnikov:2018zfn,Hubeny:2019bje,Dwivedi:2019bzh,Mironov:2019taf,Buican:2019evc,Dwivedi:2020jyx,Dwivedi:2020rlo,Fliss:2020yrd,Leigh:2021trp,Kolganov:2021mew,Dwivedi:2021dix,Nishioka:2021cxe,Bao:2021gzu,Melnikov:2022qyt,Melnikov:2023nzn,Fliss:2023dze,Fliss:2023uiv,Melnikov:2023wwc,Chung:2023gwk}. Knot theory can be implemented in many different research lines, see for instance, \cite{Aravind}. One of the most interesting applications is given by topological quantum computing, where the computations are codified in the link configurations associated with anyons \cite{Moore:1991ks,Kitaev:1997wr,Aharonov:2006yxu,Nayak:2008zza}. Recent development in this direction can be found in references \cite{Melnikov:2017bjb,Kolganov:2021mew}. An overview describing some of these advances was discussed by Melnikov in reference \cite{Melnikov:2020mno}.

In \cite{Balasubramanian:2016sro,Balasubramanian:2018por}, the multi-boundary entanglement entropy of link configurations is explored through the usual von Neumann EE by using the replica trick and it is explicitly shown the relation between the entanglement of the subsystems of links and the linking of the Wilson loops. In \cite{Balasubramanian:2016sro}, the Abelian U$(1)_k$ Chern-Simons theory was used to successfully obtain a general entanglement entropy expression for an arbitrary $n$-component link. Additionally, the non-Abelian SU$(2)_k$ Chern-Simons theory was used to analyse specific cases of link configurations of two and three-components. Moreover, it was shown that their associated entanglement entropies were sensitive not only to the linking number of the components of the links but also to more elaborated topologies of the individual components. For example, the U$(1)_k$ entanglement entropy of the Whitehead link $5^2_1$ is zero whereas its SU$(2)_k$ entanglement entropy is non-vanishing. Following \cite {Balasubramanian:2016sro,Balasubramanian:2018por}, $5^2_1$ is in the Rolfsen notation the link with 5 crossings, 2 components and 1 chronological time. Consequently, the entanglement is related to a non-trivial link, whose EE is non-vanishing. This is a realization of the Aravind's conjecture reached by the introduction of the EE of links \cite{Aravind,Melnikov:2020mno}. 

A general expression for the entanglement entropy of SU$(2)_k$ CS theory has proven challenging to derive, even for $3$-component links. Consequently, obtaining an entanglement entropy closed-form expression for $n$-component links appears unattainable.  Our manuscript aims to pursue research in this direction by explicitly computing the entanglement entropy of the connected sum of an arbitrary number of Hopf links, gaining some insight in the understanding of the entanglement in multi-partite systems for the non-Abelian CS case. Some work in this direction has been carried out for instance in Refs.  \cite{Balasubramanian:2018por,Dwivedi:2020jyx,Fliss:2020yrd}, and some general formulas are found for the case of torus links, it was not realized a similar convergent behaviour as the one discussed in the present article for Hopf links.

The structure of this paper is as follows. In section \ref{EEinCS} we give a brief overview on the computation of the EE of links in CS theory with Abelian and non-Abelian gauge groups for the connected sum of links. In section \ref{Hopf link} the calculation of the EE of the connected sum of three Hopf links is carried out. This same section contains the generalisation for the connected sum of an arbitrary amount of Hopf links. Finally, in section \ref{conclusions} we discuss the paper's results and some perspectives for future work.

\section{Entanglement entropy in Chern-Simons theory}
\label{EEinCS}

In the present section we give an brief overview of preliminaries and basic material to compute the entanglement entropy for links in the Chern-Simons gauge theory for the Abelian and non-Abelian case with group SU$(2)_k$ with level $k$. We will not be exhaustive, we only provide the minimal details to introduce notation and conventions for future reference. To be concrete we focus on the Hopf links.

We start by considering an $n$-component link $\mathcal{L}^n = L_1 \cup L_2 \cup \dots \cup L_n$ (the disjoint union of $n$ link components) in $S^3$. The associated quantum state $|\mathcal{L}^{\otimes n} \rangle = |\mathcal{L}_1 \rangle \otimes |\mathcal{L}_2 \rangle \otimes \dots \otimes |\mathcal{L}_n \rangle$ can be obtained by performing the Euclidean path integral of Chern-Simons theory on the link complement manifold specified by $M_n = S^3 \setminus \widetilde{\mathcal{L}}^n$, where $\widetilde{\mathcal{L}}^n$ is a tubular neighbourhood of $\mathcal{L}^n$ in $S^3$. This quantum state belongs to the $n$-component Hilbert space $\mathcal{H}^{\otimes n} = \mathcal{H}_1 \otimes \mathcal{H}_2 \otimes \dots \otimes \mathcal{H}_n$ associated with the boundary $\partial M_n$. This boundary is composed by $n$ tori, one for each link component. 

Consider a $3$-manifold $M$ constructed as the connected sum of manifolds $M_L$ and $M_R$, i.e., $M=M_L\#M_R$. If we have two Hopf links $\mathcal{L}_1$ in $M_L$ and $\mathcal{L}_2$ in $M_R$, it is proven in \cite{Witten:1988hf} that the normalized total partition function is given by
\begin{equation}
\frac{Z(M,\mathcal{L})}{Z(M)} = \frac{Z(M_L,\mathcal{L}_1)}{Z(M)} \cdot \frac{Z(M_R,\mathcal{L}_2)}{Z(M)}, 
\end{equation}
where $\mathcal{L}_1 \# \mathcal{L}_2$ is the connected sum of the links $\mathcal{L}_1$ and $\mathcal{L}_2$. This expression can be used to obtain the partition function of a three-component Hopf link in $S^3$. This construction admits a generalization applied to the connected sum of $r-1$ Hopf links $\mathcal{L} = \mathcal{L}_1 \# \cdots \# \mathcal{L}_{r-1}$ with $r$ components. The total partition function is the product of $r$ normalized factors
\begin{equation}
\frac{Z(M,\mathcal{L})}{Z(M)} = \prod_{i=1}^{r-1} \frac{Z(M_i,\mathcal{L}_i)}{Z(M)},
\end{equation}
which allows us to construct multi-partite systems easy to analyse in CS theory.
The entanglement entropy associated with the bi-partition $(L_1,\dots, L_m | L_{m+1},\dots, L_n)$ of $\mathcal{L}^n $ into two sub-links $\mathcal{L}^m = L_1 \cup \dots \cup L_m$ and $\mathcal{L}^{n-m} = L_{m+1} \cup \dots \cup L_n$ is given by the von Neumann entropy
\begin{align}
S_{EE (\mathcal{L}^m | \mathcal{L}^{n-m})} (\mathcal{L}^n) =
- \mathrm{Tr}_{L_{m+1}, \dots, L_n} \left( \rho_{m+1, \dots, n} \ln  \rho_{m+1, \dots, n} \right),
\label{linkentropy}
\end{align}
where the reduced density matrix $\rho_{m+1, \dots, n}$ associated with the sub-link $\mathcal{L}^{n-m}$ is given by
\begin{align}
\rho_{m+1, \dots, n} =
\frac{1}{\langle \mathcal{L}^n | \mathcal{L}^n \rangle}
\mathrm{Tr}_{L_1, \dots, L_m} 
| \mathcal{L}^n \rangle \langle \mathcal{L}^n |.
\label{reduceddensitymatrix}
\end{align}

In \cite{Balasubramanian:2016sro}, the U$(1)_{k}$ Chern-Simons theory was used to provide a general expression for the EE associated with an arbitrary bi-partition of $\mathcal{L}^n$. The EE was found to be
\begin{align}
S_{EE (\mathcal{L}^m | \mathcal{L}^{n-m})} (\mathcal{L}^n) = \ln \left( \frac{k^m}{|\textrm{ker}(\mathbf{G})|} \right),
\label{SEEabelian}
\end{align}
where $\mathbf{G}$ is the so-called linking matrix across the bi-partition, and it is written as
\begin{align}
\mathbf{G}=
\begin{pmatrix}
\it{l}_{1,m+1} & \it{l}_{2,m+1} & \cdots & \it{l}_{m,m+1} \\
\it{l}_{1,m+2} & \it{l}_{2,m+2} & \cdots & \it{l}_{m,m+2} \\
\vdots & \vdots & & \vdots \\
\it{l}_{1,n} & \it{l}_{2,n} & \cdots & \it{l}_{m,n}
\end{pmatrix},
\label{linkingmatrix}
\end{align}
where $\it{l}_{i,j}$ is the linking number modulo $k$ between $L_i$ and $L_j$. It was also proven in reference \cite{Balasubramanian:2016sro} that the entanglement entropy (\ref{SEEabelian}) vanishes if and only if all the elements of $\mathbf{G}$ are zero modulo $k$.
Thus, the entanglement entropy of a U$(1)_{k}$ Chern-Simons theory is sensitive to the link components' linking number modulo $k$.

In $SU(2)_k$ Chern-Simons, the entanglement entropy can be computed for an arbitrary link $\mathcal{L}^n$ within U$(1)_{k}$ Chern-Simons theory. However, it has only been computed for some particular cases in SU$(2)_{k}$ Chern-Simons theory \cite{Balasubramanian:2016sro}.
In this section, the connected sum of an arbitrary number of Hopf links is studied. 

\section{Entanglement entropy of Hopf links}
\label{Hopf link}

We begin by analysing the Hopf link $2^2_1$, which is the simplest non-trivial two-component link. For the gauge group SU$(2)_k$, this link is maximally entangled and it is associated with the quantum state
\begin{align}\label{Hopf quantum state}
    \big| 2^2_1 \big\rangle = \sum_{j_1, j_2} \mathcal{S}_{j_1 j_2} |j_1,j_2\rangle,
\end{align}
where the indices $j_1$ and $j_2$ run over all the integrable representations of SU$(2)_k$, so the $|j_1,j_2\rangle$ quantum states form a basis of the Hilbert space. Additionally, the unitary transformation $\mathcal{S}$ is expressed by \cite{Witten:1988hf}

\begin{align}
    \mathcal{S}_{j_1 j_2}=\sqrt{\frac{2}{k+2}} \sin{\left(\frac{(2 j_1 + 1)(2 j_2 +1) \pi}{k+2}\right)}.
\label{deltarelation}
\end{align}

To obtain the entanglement entropy of the Hopf link, we need to compute the reduced normalized density matrix (\ref{reduceddensitymatrix}). By tracing out the first component, we obtain
\begin{align}
    \rho_2(2^2_1)=\frac{1}{\langle 2^2_1|2^2_1\rangle} \mathrm{Tr}_{L_1}  \big|2^2_1\rangle\langle 2^2_1\big| = \frac{1}{\mathrm{dim}\,\mathcal{H}(T^2)}\sum_j \big|j\rangle\langle j\big|.
\end{align}
This expression is proportional to the identity matrix, which implies the state is maximally entangled. Finally, the entanglement entropy can be computed by using \eqref{linkentropy} and is given by
\begin{align}
    S_{EE}(2^2_1)= \ln\mathrm{dim}\,\mathcal{H}(T^2) = \ln (k+1),
\label{EE2-components}
\end{align}
which is in complete agreement with the result obtained in reference \cite{Balasubramanian:2016sro}.

\subsection{Connected sum of two Hopf links}
\label{2Hopf}

The prescription in the seminal paper \cite{Witten:1988hf} for computing the partition function of the connected sum of two links can be summarised in the following expression
\begin{align}\label{connected-general}
    \frac{Z \left( S^3; \mathcal{L}_1 \# \mathcal{L}_2 \right)}{ Z \left( S^3; C \right)} = \frac{Z \left( S^3; \mathcal{L}_1 \right)}{ Z \left( S^3; C \right)} \cdot \frac{Z \left( S^3; \mathcal{L}_2 \right)}{ Z \left( S^3; C \right)},
\end{align}
where $C$ is the strand where the surgery is performed. This formula can be used to construct the corresponding quantum states of the connected sum configurations. For the connected sum of two Hopf links, it is written as follows
\begin{align}
\label{connected-2Hf}
\frac{Z\left(L(R_1,R_2,R_3)\right)}{Z\left(R_2\right)} = \frac{Z\left(L(R_1,R_2)\right)}{Z\left(R_2\right)} \cdot \frac{Z\left(L(R_2,R_3)\right)}{Z\left(R_2\right)},
\end{align}
where we have omitted the $S^3$ from the formula. 

The corresponding quantum state for the connected sum of two Hopf links can be obtained from expression \eqref{connected-2Hf} and is given by
\begin{align}
\big| 2^2_1 + 2^2_1 \big\rangle 
= \sum_{j_1, j_2, j_3} 
\frac{\mathcal{S}_{j_1 j_2} \mathcal{S}_{j_2 j_3}}{\mathcal{S}_{0 j_2}}
\big| j_1, j_2, j_3 \big\rangle,
\label{stateN=4}
\end{align}
where the sum is over all the integrable representations of SU$(2)$ at level $k$. For this case, we have three possible bi-partitions, namely, $(L_1|L_2,L_3)$, $(L_2|L_1,L_3)$, and $(L_3|L_1,L_2)$. We are interested in the behaviour of the entanglement entropy as the number of link components increases. For this reason, we will focus on bi-partitions of the form $(L_1|L_2,\dots,L_n)$, where $n$ is the number of link components. For the connected sum of two Hopf links, such bi-partition is $(L_1|L_2,L_3)$. The reduced normalised density matrix for this bi-partition is
\begin{align}\label{density-2Hf}
    \rho_{2,3}=\frac{1}{\mathcal{N}}\sum_{l,i,j} \frac{\mathcal{S}_{il}\mathcal{S}_{lj}}{|\mathcal{S}_{0l}|^2}\ |l,i\rangle\langle l,j|,
\end{align}
with the normalisation factor $\mathcal{N}=\big\langle 2^2_1 + 2^2_1 \ \big| \ 2^2_1 + 2^2_1 \big\rangle = \sum_{k} |\mathcal{S}_{0k}|^{-2}$.
For computing the entanglement entropy, it is convenient to use the replica method
\begin{align}\label{replicamethod}
    S_{EE(\mathcal{L}^m | \mathcal{L}^{n-m})} (\mathcal{L}^n) = - \frac{d}{d \alpha} \bigg(\mathrm{Tr}_{L_{m+1}, \dots, L_n} \left(\rho_{m+1, \dots, n} \right)^\alpha\bigg)\bigg|_{\alpha=1}.
\end{align}
Using expression \eqref{replicamethod}, the EE of the connected sum of two Hopf links can be obtained; it is given by
\begin{align}\label{EE-3components}
    S_{EE} \big(2^2_1 + 2^2_1\big) = -\sum_{j} p_{j} \ln p_{j}, 
   \end{align}
where
\begin{equation}
 p_j = \frac{\mathcal{S}_{0j}^{-2}}{\sum_l \mathcal{S}_{0l}^{-2}}.
\end{equation}
It is possible to argue that $S_{EE}$ is independent of the chosen bi-partition \cite{Balasubramanian:2016sro}.

In figure \ref{fig:2,3-Hopf}, the entanglement entropy of the Hopf link \eqref{EE2-components} and the connected sum of two Hopf links \eqref{EE-3components} is compared as a function of $k$. As expected, the EE of the Hopf link ($2$-component curve) is the upper bound for the one associated with the connected sum of two Hopf links ($3$-component curve). This behaviour is consistent with the characterisation of the Hopf link as a maximally entangled configuration. In the next section, we will show evidence that the $3$-component curve in figure \ref{fig:2,3-Hopf} is the lower bound for the family of curves with an arbitrary number of components.

 \begin{figure}[ht]
            \centering
            \includegraphics[scale=0.60]{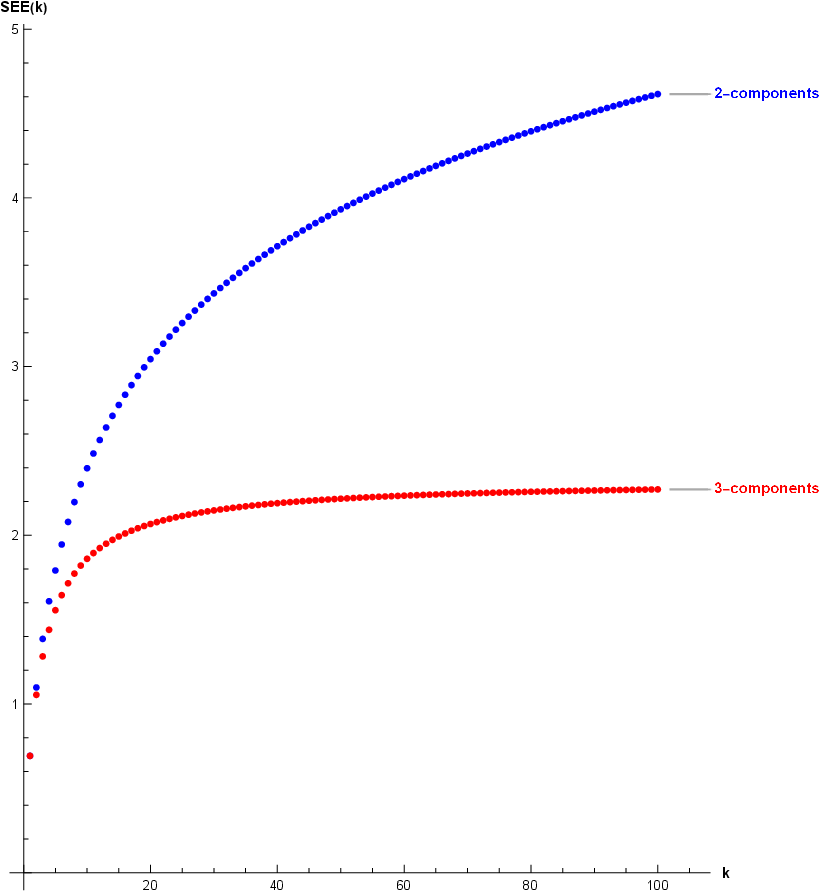}
                \caption{Entanglement entropy of the Hopf link (blue) and the connected sum of two Hopf links (red) as a function of $k$.}
            \label{fig:2,3-Hopf}
        \end{figure}

\subsection{Connected sum of an arbitrary number of Hopf links}
\label{NHopf}

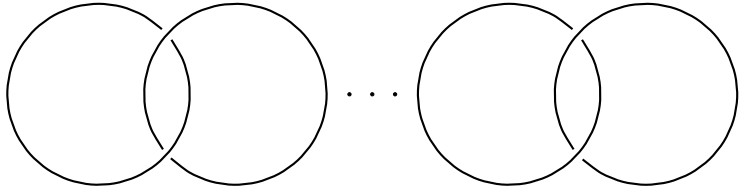
\begin{figure}[ht]
   \centering
   \begin{tikzpicture}[scale = 0.6]
      \draw[domain=46:397,thick] plot[smooth] ({-6+2*cos(\x)}, {2*sin(\x)});
        \draw[domain=225:577.5,thick] plot[smooth] ({-3+2*cos(\x)}, {2*sin(\x)});
        \filldraw (-0.5,0) circle (1pt);
        \filldraw (0,0) circle (1pt);
        \filldraw (0.5,0) circle (1pt);
        \draw[domain=46:397,thick] plot[smooth] ({3+2*cos(\x)}, {2*sin(\x)}); 
        \draw[domain=226:577.5,thick] plot[smooth] ({6+2*cos(\x)}, {2*sin(\x)});
    \end{tikzpicture}
    \caption{Connected sum of $n-1$ Hopf links.}
    \label{fig:n-hopf}
\end{figure}

In this section, our main result was obtained: the entanglement entropy of the connected sum of $n-1$ Hopf links for the bi-partition $(L_1|L_2,\dots,L_n)$ was computed and its behaviour as the number of components changes was explored.

The $n$-component link $L_1\cup\cdots\cup L_n$ is constructed as the connected sum of an ($n-1$)-component link and a Hopf link. Thus the expression \eqref{connected-general} for this case is
\begin{align}
\frac{Z\left(L(R_1,R_2,\dots,R_{n})\right)}{Z\left(R_2\right)} = \frac{Z\left(L(R_1,R_2)\right)}{Z\left(R_2\right)} \cdot \frac{Z\left(L(R_2,\dots,R_{n})\right)}{Z\left(R_2\right)},
\end{align}
which can be further decomposed into smaller links as follows:
\begin{align}
Z(L(R_1,R_2,\dots,R_n)) = \frac{Z(L(R_1,R_2))}{Z(R_2)} \cdot \frac{Z(L(R_2,R_3))}{Z(R_3)} \cdots \frac{Z(L(R_{n-1},R_n))}{Z(R_{n-1})}.
\label{connected-N}
\end{align}

By using \eqref{connected-N} we can obtain the quantum state of the connected sum of $n-1$ Hopf links; it can be written as
\begin{align}\label{N-Hf state}
    \big|2^2_1+2^2_1+\dots+2^2_1\big\rangle = \sum_{J}\mathcal{S}_{J}|J\rangle,
\end{align}
where we have introduced the following notation
\begin{align}
    \sum_{J}\mathcal{S}_{J}|J\rangle= \sum_{j_1,\dots,j_n}\frac{\mathcal{S}_{j_1j_2}\dots\mathcal{S}_{j_{n-1}j_{n}}}{\mathcal{S}_{0j_2}\dots\mathcal{S}_{0j_{n-1}}}|j_1,\dots, j_n\rangle.
\end{align}
To get the entanglement entropy, we need the density matrix associated with the state \eqref{N-Hf state}, which is obtained by performing the exterior product of the quantum state with itself; we get
\begin{equation}
    \rho = \frac{1}{\mathcal{N}}\sum_{J}\sum_{K}\mathcal{S}_{J}\,\mathcal{S}^*_{K}\,|J\rangle\langle K|,
\end{equation}
where
\begin{equation}
\mathcal{N} = \sum_J |\mathcal{S}_J|^2.
\end{equation}
As established before, we are interested in the system's behaviour as the number of components increases. For this reason, we restrict to the bi-partition $(L_1|L_2,...,L_n)$. The reduced density matrix of the sub-link $L_2\cup\cdots\cup L_n$ is obtained by tracing out the first component, and is given by
\begin{align}
    \rho_{L_2,...,L_n}=\frac{1}{\mathcal{N}}\sum_{J-j_1}\sum_{K-k_1}\mathcal{S}_{J-j_1}\delta_{k_2 j_2}\mathcal{S}^*_{K-k_1}|J-j_1\rangle\langle K-k_1|.
\end{align}

It is straightforward to compute the $n$-power of the density matrix, we get
\begin{align}
    \rho^n_{L_2,...,L_n}=\frac{1}{\mathcal{N}^n}\sum_{J-j_1}\sum_{M-m_1}\mathcal{S}_{J-j_1}\delta_{j_2 m_2}\mathcal{S}_{M-m_1}^*\left(\sum_{K}\left|\mathcal{S}_{K}\right|^2 \delta_{j_2 k_2}\right)^{n-1} |J-j_1\rangle\langle M-m_1|.
\end{align}
By tracing over the one-dimensional Hilbert spaces associated with the components $L_2$,\dots,$L_n$, we obtain
\begin{align}\label{anterior}
    \mathrm{Tr}_{L_2,\dots, L_n}\,\rho^n_{L_2,...,L_n}=\frac{1}{\mathcal{N}^n}\sum_{j_2}\left(\sum_{K}\left|\mathcal{S}_{K}\right|^2 \delta_{j_2 k_2}\right)^n,
\end{align}
and we use expression \eqref{replicamethod} to compute the entanglement entropy 
$$
S_{EE(L_1 |L_2,...,L_n)}(2^2_1+2^2_1+\dots+2^2_1)=-\frac{d}{dn} \bigg( \mathrm{Tr}_{L_2,\dots,L_n}\ \rho^n_{L_2,...,L_n} \bigg) \bigg|_{n=1}
$$
\begin{equation}\label{SEE-N}    
    =-\sum_j p_j \ln p_j,
\end{equation}
where we have introduced the quantity
\begin{align}
    p_{j}=\frac{1}{\mathcal{N}}\sum_{K}\left|\mathcal{S}_{K}\right|^2 \delta_{j k_2}.
\end{align}

\begin{figure}[ht]
            \centering
            \includegraphics[scale=0.60]{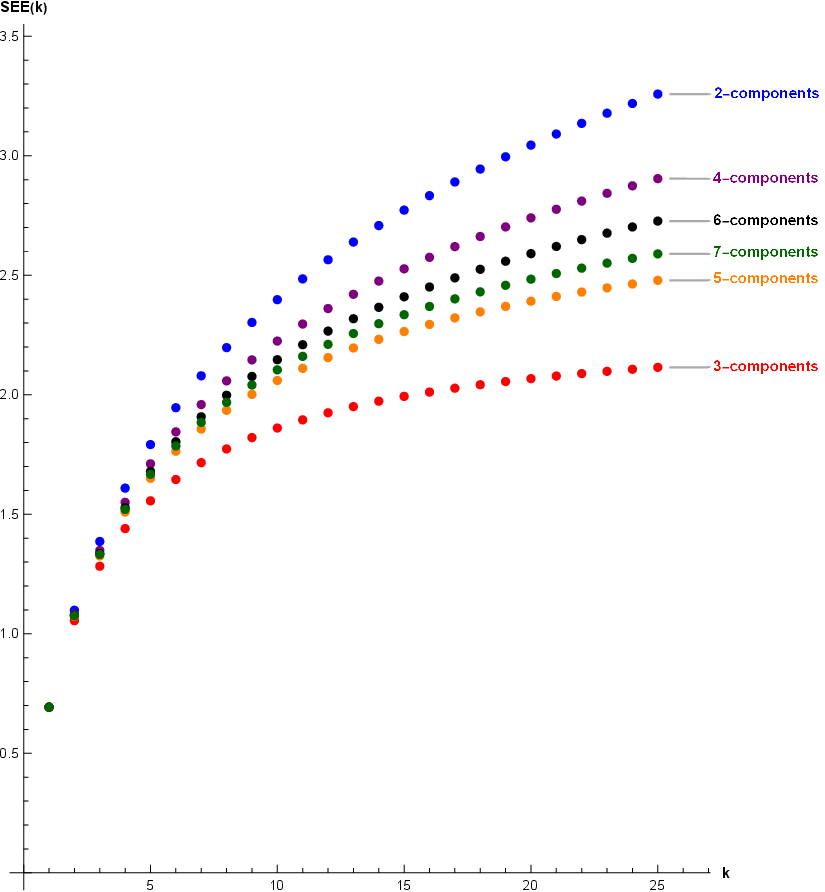}
                \caption{In this figure it is shown a family of curves which are bounded from above and below. The upper bound corresponds to the curve of the maximally entangled Hopf link. It is also displayed how the curve associated with the connected sum of two Hopf links is actually a lower bound. It is also shown the emergence of a convergent behaviour of the curve collection. It is seen how curves with an even number of components approach the limit from above and those with an odd number converge from below.}
            \label{Plot2-7}
        \end{figure}

Once equation \eqref{SEE-N} has been obtained, we can analyse the behaviour of the entanglement entropy as a function of $k$ for different number of components $n$. The characterisation of the EE can be extracted from the curves associated with links having a few number of components, this is shown in figure \ref{Plot2-7}. We can observe that the family of curves is bounded from above and below. As expected, the upper bound corresponds to the curve of the maximally entangled Hopf link. Furthermore, we found that the curve associated with the connected sum of two Hopf links is actually a lower bound. The most striking and unexpected result from this plot is the emergence of a convergent behaviour of the curve collection, with the peculiarity curves with an even number of components approach the limit from above, whereas those with an odd number converge from below.

\section{Conclusions}
\label{conclusions}

In this article, we find the entanglement entropy of the connected sum of Hopf links with an arbitrary number of components.  We have obtained a closed form for the entanglement entropy restricted to the bi-partitions of the form $(L_1|L_2,\dots,L_n)$. We have shown that Chern-Simons theory with group SU$(2)$ is sensitive to the parity of the links, that means it recognises when a link has an even or odd number of components. Additionally, we have also found evidence of the existence of a limit curve as the number of components increases, $n\to\infty$.

It will be interesting to investigate the behaviour of the entanglement entropy as non-trivial knot components are introduced into the link. Another venue will be to explore the properties captured by the EE when we identify the first and the last components of the link, ending in a necklace link.

The entanglement entropy \eqref{SEE-N} associated with the connected sum of Hopf links is presented in figure \ref{Plot2-7} for link configurations from $2$ to $7$ components and for $k$ an integer in the interval $[1,25]$. The construction was performed in Mathematica, one code for each configuration. It is worth mentioning that the Mathematica codes can be easily extended to higher number of components, but the processing time grows rapidly as $n$ increases.

Finally, it is worth pointing out that the computation of the EE of links for a non-orientable RCFT might be carried out following the prescriptions of  Refs. \cite{Garcia-Compean:2018ury,Horava:1990ee}. The EE of links worked out in the present paper could be related to other kind of entropies such as the topological and crosscaps entropies described in these references.

\vskip 1 truecm
\centerline{\bf Acknowledgements}

This work was partially supported by ``Secretaría de Educación, Ciencia, Tecnología e Innovación de la Ciudad de México (SECTEI)" of Mexico City, by CONAHCyT via the SNI research assistant program, and by the Leinweber Center for Theoretical Physics (LCTP), Randall Laboratory of Physics, The University of Michigan.

\end{document}